%% file: ms.tex
\documentclass[12pt,preprint]{aastex}

\title{Parameterization studies of the properties of the X-ray dips for Low Mass X-ray binary
X1916-053}
\author{Chin-Ping Hu, Yi Chou, and Yi-Ying Chung}
\affil{Graduate Institute of Astronomy, National Central
University, Chung-Li 32054, Taiwan, ROC}
\email{m929011@astro.ncu.edu.tw, yichou@astro.ncu.edu.tw,
m939012@astro.ncu.edu.tw}

\begin{document}

\begin{abstract}

The ultra-compact Low Mass X-ray Binary (LMXB) X1916-053, composed of a neutron star and a semi-degenerated white dwarf, exhibits periodic X-ray dips with variable width and depth. We have developed new methods to parameterize the dip to systematically study its variations. This helps to further understand binary and accretion disk behaviors.  The RXTE 1998 observations clearly show a 4.87d periodic variation of the dip width. This is probably due to the nodal precession of the accretion disk, although there are no significant sidebands in the spectrum from the epoch folding search. From the negative superhump model \citep{larwood}, the mass ratio can be estimated as $q=0.045$. Combined with more than 24 years of historical data, we found an orbital period derivative of $\dot{P}_{orb}/P_{orb}=(1.62 \pm 0.48)\times 10^{-7} yr^{-1}$ and established a quadratic ephemeris for the X-ray dips.  The period derivative seems inconsistent with the prediction of the standard model of binary orbital evolution proposed by \citet{rappaport1987}. On the other hand, the radiation-driven model \citep{tavani1991} may properly interpret the period derivative even though the large mass outflow predicted by this model has never been observed in this system.  With the best ephemeris, we obtained that the standard deviation of primary dips are smaller than that of secondary dips.  This means that the primary dips are more stable than the secondary dips.  Thus, we conclude that the primary dips of X1916-053 occur from the bulge at the rim instead of the ring of the disk proposed by \citet{frank1987}.
\end{abstract}

\keywords{stars: individual (X1916-053) --- X-rays: binaries ---
X-rays: individual (X1916-053)}

\section{Introduction}

Some X-ray binaries show periodic dips in their X-ray light curves.  To date, there are eleven X-ray binaries that known for periodic dipping features \citep{ritter2003}.  It is widely believed that these X-ray dips are caused by X-rays from the region around the compact object being periodically absorbed by the vertical structure on the outer part of accretion disk.  The periodic dips provide us an opportunity to measure the orbital period and investigate the accretion disk dynamics of the binary system.  

The low-mass X-ray binary X1916-053 (4U1915-05), composed of a neutron star and a white dwarf companion, shows various timing properties.  For the binary orbital period, it can be obtained by periodic X-ray intensity dips in light curves.  \citet{walter1982} and \citet{white1982} first discovered its X-ray dips with Einstein and OSO-8 observations.  \citet{white1982} found two possible periods of $49.93\pm 0.06$min and $50.06\pm 0.03$min, which is the shortest one among all X-ray dippers.  \citet{homer2001} used the O-C (Observed - Calculated) technique on the 1998 RXTE observation data to calculate the best fitting period of $3000.6\pm 0.2$s.  \citet{chou2001} used 1979-1996 data to derive the orbital period of $3000.6508\pm 0.0009$s and established an X-ray dip ephemeris.  In addition to the $\sim$3000s recurrent dips, secondary dips are also occasionally seen at phases from $\sim 0.4$ to $\sim 0.6$ relative to the center of primary dips.

The optical modulation period of X1916-053 is 3027.5510 $\pm$ 0.0052 (s), which is only $\sim$ 1\% significantly longer than the X-ray dip period \citep{chou2001}.  Optical observations have revealed the 3000s X-ray period \citep{grindlay1989, grindlay1992}, whereas the 3028s optical period with a series of $\sim$3.9d side bands was detected in X-ray light curves \citep{chou2001}.  The optical period has ones been considered as the orbital period of the system and the X-ray dip period is the beat period of binary orbital period and a third companion with orbital period $\sim$3.9d \citep{grindlay1988}.  However, \citet{chou2001} concluded that the X-ray dip period, like other dipping sources, is the orbital period. Further, they concluded that the optical modulation is likely caused by the orbital motion coupled with a 3.9d disk apsidal precession period as the superhump in SU UMa type dwarf novae.  Furthermore, \citet{retter} discovered a negative superhump with a period of 2979.3s , which is the beat period of orbital period and $\sim$4.8d  disk nodal precession period.  

On the other hand, dip shape modulation was first found in Ginga data  \citep{yoshida}.  Due to short observation duration, the modulation period could only be constrained to $\sim$ 5-6 days.  \citet{chou2001} found that the dip phase probably modulated within a 4.85d or 3.76d period by analyzing 10-day consecutive RXTE observation in 1996.  A dip phase variation with amplitude $\sim$0.05 and period of 4.74d was seen in 10-day consecutive RXTE observations in 1998 \citep{homer2001}.   

Dip shape modulation provides an opportunity to study the dynamics of the disk. Unfortunately, dip shape has never been quantitatively investigated before.  For example, in order to investigate the possible periods between 3.5 and 5d, \citet{homer2001} have considered the area of dip, length of dip, depth of dip, and phase of deepest dip, but none of them have clear indication of $\sim$4d cycle.  In this paper, we have developed a new method to parameterize the properties of the X-ray dip to further investigate its variation using archived data from various observatories with a total time span of more than 24 years.

In section \ref{od}, we briefly introduce the historical data applied in this study and the corresponding basic reductions. Section \ref{da} describes data analysis methods, including the definition of the dip center, width and strength, as well as the results, such as the periodic variation of dip width, the long-term orbital ephemeris and the stabilities of primary and secondary dips.  Finally, we discuss interpretations from our analysis of the nodal precession of accretion disk and the period derivative of the system in Section \ref{discuss}.

\section{Observation and Data Reduction}\label{od}

Many X-ray observatories have observed X1916-053 since its discovery. More than 24 years of observation data, from OSO-8 (1978) to XMM-Newton (2002), have been reduced through standard processes and archived on the HEASARC (NASA's High Energy Astrophysics Science Archive Research Center) website. Table \ref{obs_status} lists the historical observations of X1916-053 used in this paper. RXTE performed two 10-day consecutive observations in 1996 and 1998, which enable us to study dip parameter modulations over a time scale of several days.  Simultaneous observations by RXTE and XMM-Newton were also proceeded in May 2002.

X1916-053 was observed by XMM-Newton on 2002 September 25 for about 15000 seconds. The XMM-Newton data were reduced through the standard process with SAS (XMM-Newton Science Analysis System). All the instruments, including MOS detector, pn detector, RGS spectrometer, and OM module, were used for this observation. Although pn detector provided larger count rate and better signal to noise ratio, it only detected four complete dips while MOS detector observed five dips.  In order to collect more number of dips for better statistics, we used MOS data for dip phase analysis.  The barycenter correction, which corrects the time system from satellite to the barycenter of the solar system, is important for the timing analysis of such a short orbital period. SAS provided the barycenter correction subpackage, {\it barycen}, for the XMM-Newton data. After applying the
barycenter correction, the data can then be used for further dip analysis.

X1916-053 was observed by ASCA on 1993 May 2 for about 67ks and by ROSAT on 1992 October 17 for about 18ks. Since FTOOLS
provided the tools for correcting the time system, we downloaded the event files for ASCA and ROSAT. The ASCA data barycenter correction was produced by the subpackage {\it timeconv} while the ROSAT data used {\it bct+abc}. In addition, the event files of BeppoSAX observations were included in our analysis. The observation time was approximately 82ks in 1997 and 98ks in 2001. The data were barycenter corrected with the FTOOLS subpackage {\it earth2sun}, which can only correct the effect of earth to the solar system barycenter.

The data collected by RXTE, Ginga, EXOSAT, Einstein, and OSO-8
were downloaded from HEASARC in light curve format already reduced through corresponding standard processes.Additional column BARYTIME in RXTE light curve data contains barycenter corrected time system by FTOOLS subpackage {\it fxbary}.  All the other data sets were barycentric corrected by {\it earth2sun} only.

\section{Data Analysis}\label{da}

\subsection{Dip Parameters}

\begin{figure}
\plotone{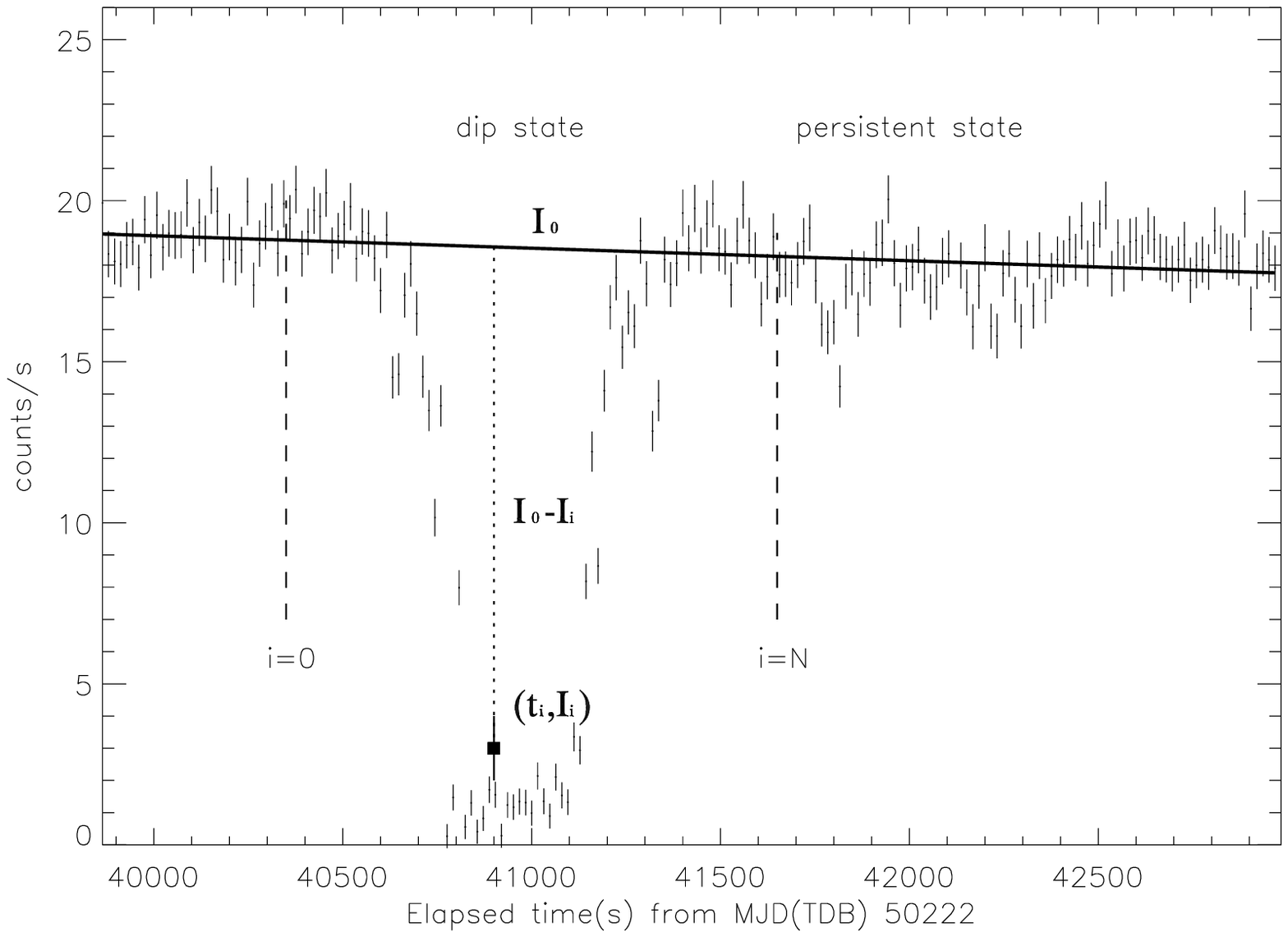} \caption{Definition of dip center times. This is an example light curve taken by RXTE in 1996.  Where $I_0$ is the persistent state fitting result, $I_i$ is the observed count rate, and $t_i$  is the corresponding time. $I_0-I_i$ is the weighting factor in our definition of dip parameters.} \label{definition}
\end{figure}

Researchers have long been aware of variations of dip phase and width, as well as other parameters \citep{yoshida,chou2001,homer2001}.  Unfortunately, dip profiles
are usually complicated, and finding a unique model of parameters applicable to all dips is difficult. Taking the definition of a dip center as an example, \citet{yoshida} defined dip center time as the midpoint of selected ``start'' and ``end'' times of a dip, but the value highly depends on how to choose the ``start'' and ``end'' times of a dip. On the other hand, \citet{chou2001} defined dip center time by fitting a dip with a quadratic curve around the point with minimum intensity.  \citet{homer2001} fitted the dip with a Gaussian profile. However, none of these methods are suitable for dips with complex profiles, such as Fig. \ref{definition}. It is therefore necessary to extract dip properties with a new method that is not only independent of dip boundary selection but which can also be applied to all kinds of dips profiles, regardless of their complexity.

An X-ray dip is caused by the absorption of the bulge in the accretion disk.  We can first divide a light curve into dip states and persistent (non-dip) states by roughly guessing the boundaries of dips (see Fig.\ref{definition}).  The observed count rate during the persistent state $I_0$ when the bulge is far from the line of sight is estimated by fitting a straight line of neighboring persistent states. Much like calculating the gravitational center of an object, dip center time (or phase in the folded light curve) is the average time of the dip weighted by the difference of the dip state and the predicted persistent count rates:

\begin{equation}
t_c=\frac{\displaystyle \sum_{i=1}^N (I_0-I_i) \cdot t_i}{\displaystyle \sum_{i=1}^N (I_0-I_i)}
\end{equation}
where the $t_i$ and $I_i$ are the time and dip count rate of
$i^{th}$ bin, respectively.  To evaluate the physical width of the dip, the dispersion is defined as:
\begin{equation}
W=\sqrt{\frac{\displaystyle \sum_{i=1}^N (I_0-I_i)\cdot (t_i-t_c)^2}{\displaystyle \sum_{i=1}^N
(I_0-I_i)}}
\end{equation}

\noindent which is equal to the second moment of the dip state.

\noindent Analogous to the equivalent width in spectrographic
analysis, the equivalent width of the dip, which can also be
considered the strength of dip, is
\begin{equation}
EW=\sum_{i=1}^N \frac{I_0-I_i}{I_0} \cdot \Delta t
\end{equation}

\begin{figure}
\plotone{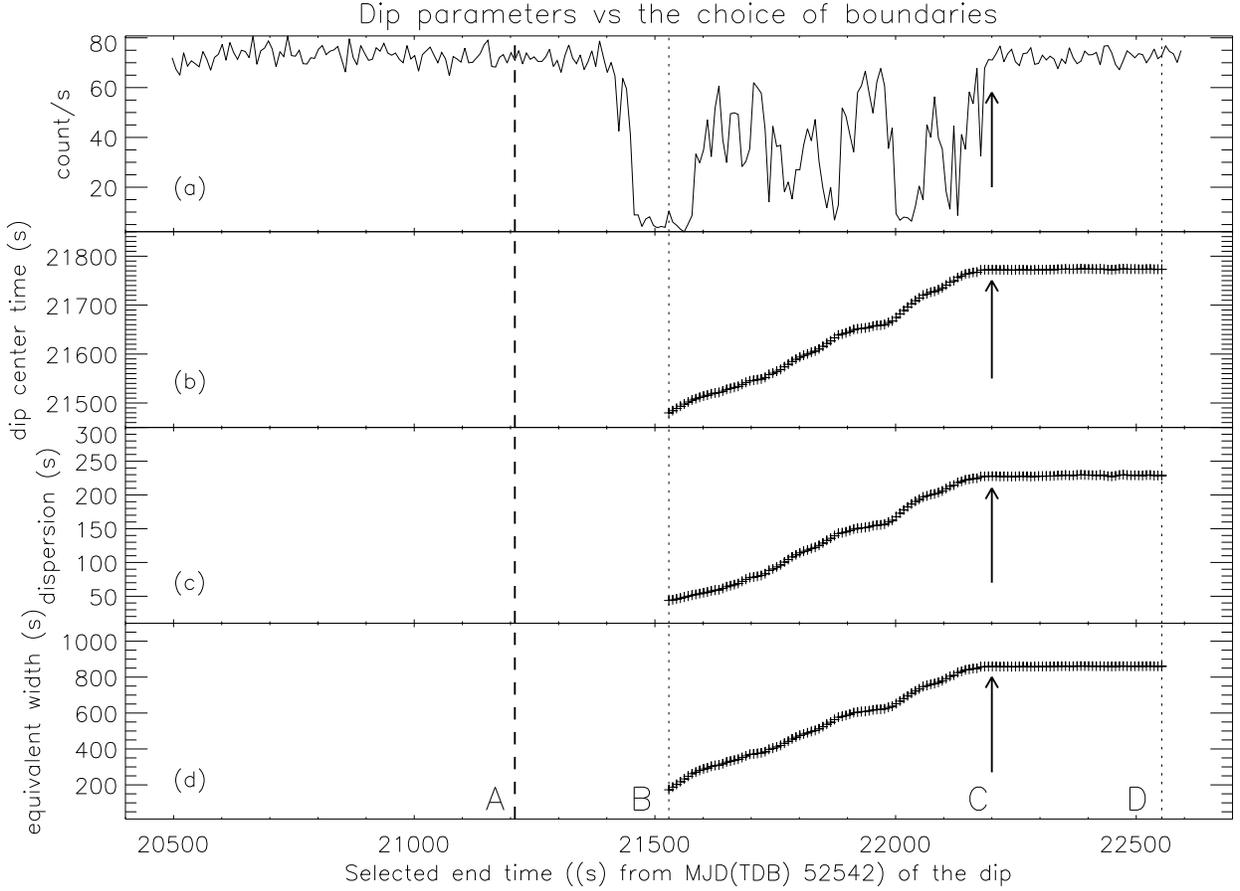} \caption{A test to show the dependence of dip parameters and light curve boundaries for evaluation.  (a) The light curve for the test.  A dip with complicated profile appeared between $\sim$21300s and $\sim$22200s (from MJD(TDB) 52542).  We set the left boundary fix at point A (21200s, the dashed line) and the right boundary (dotted line) gradually moved from point B to D.  Plot (b) (c) and (d) are the dip parameters calculated with different right boundary.  It is evidentially that as long as the right boundary lies on persistent state (beyond point C), the calculated parameters are insensitive to the location of right boundary.  }
\label{dct_boundary}
\end{figure}

By the definitions above, as long as the boundaries reside in the persistent states neighboring the dip, the points which lay on the persistent state beside the dip make little contribution to the dip parameters ($I_0-I_i \approx 0$). The parameters are therefore insensitive to boundary selection for calculations. Furthermore, these three values are well-defined regardless of the dip profile complexity. This method can also be applied to the eclipse source.  The only constraint is that a completely observable dip or eclipse is required.  In order to check if the parameters are sensitive to the boundary choice, a test was performed to a dip which is hard to fit with a quadratic function \citep{chou2001} or a Gaussian \citep{homer2001} as shown in Figure \ref{dct_boundary}.  We set the left boundary (i=1, see Figure \ref{definition}) at a fixed value and gradually change the right boundary (i=N) from dip state to persistent state.  The test result showed that the derived parameters are insensitive to the right boundary as long it well resides the persistent state.  Similar result was obtained with a fixed right boundary at persistent state and a variable left boundary.  We therefore conclude that our method is adoptable to dips of various profiles.

Simultaneous XMM-Newton and RXTE observations provide an opportunity to inspect whether the dip parameters are energy (or instrument) dependent. To test energy dependency, the XMM-Newton events detected by pn detector were divided into to a soft band (0.7-1.7 keV) and a hard band (1.7-8.0 keV) by the counts available roughly equally.  Table \ref{parameter_band} lists the parameters of four dips in pn observation. There is no significant energy dependency in dip dispersion, but the equivalent width is highly energy dependent as expected since the dip is caused by absorption.  The dip center time seems to have little energy dependency. This test was also performed with RXTE PCA data (2-9 keV) and XMM-Newton simultaneous observations. Dispersions measured from both instruments are almost the same, but the energy dependency of dip center times is still present  (see Table \ref{dct_satellite}).  The difference is apparently due to the asymmetric bulge structure with different optical depths in different energy bands. Although dip center times are energy dependent, the differences are only $\sim$ 0.01 cycle, much smaller than the phase jitter ($\sim$ 0.05 cycle, \citet{chou2001}) and can be neglected in long-term analysis.

\subsection{Periodicity of Dip Parameters}\label{pp}

\begin{figure}
\plotone{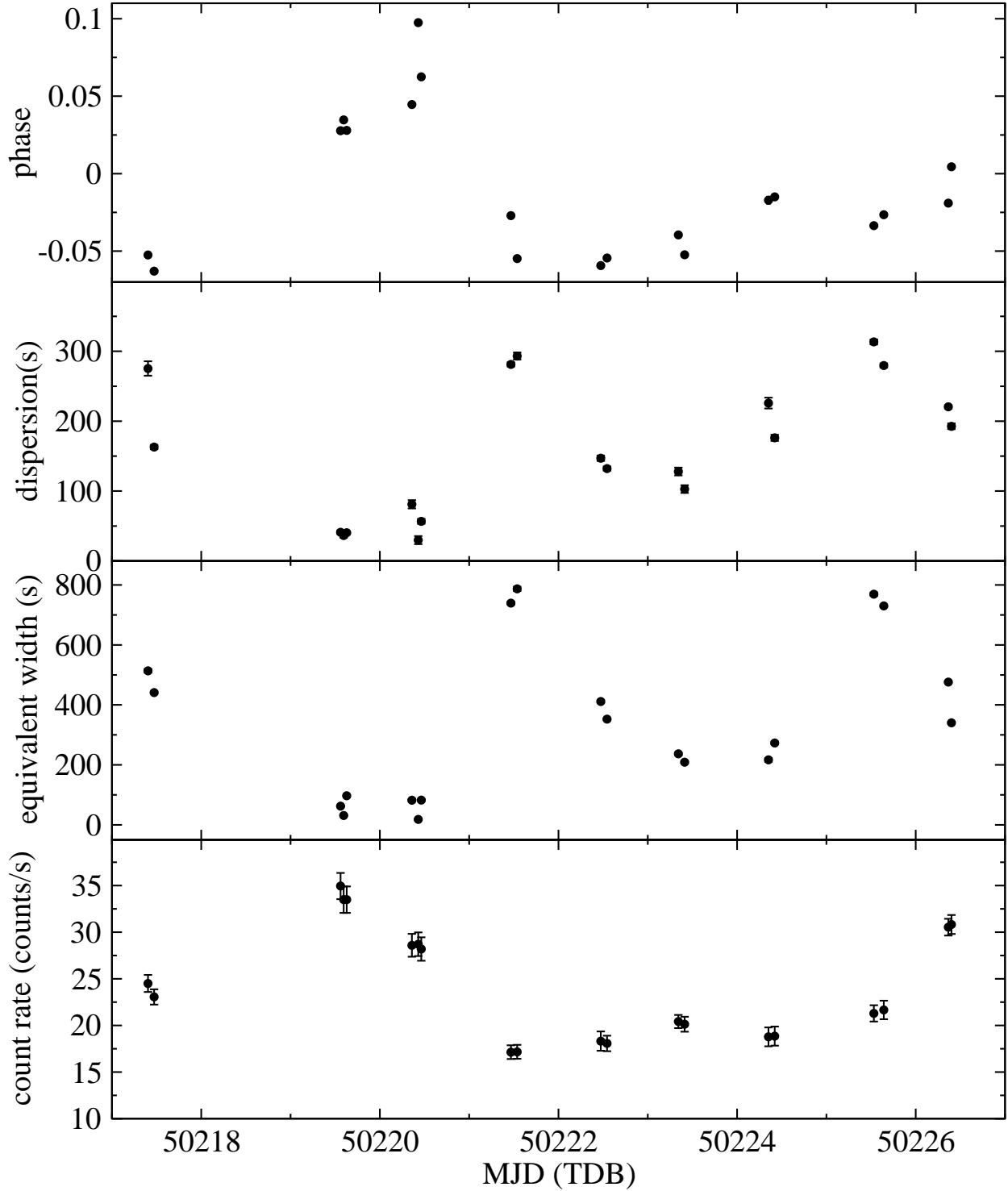} \caption{The dip parameters in 1996 10-day
consecutive observations. The first three panels represent the three dip parameters of dip with unit in seconds,  while the last panel represents the persistent count rate near each dip.  All the parameters varied over a period
of several days, but significant periodicity cannot be detected. }
\label{1996_parameter}
\end{figure}

\begin{figure}
\plotone{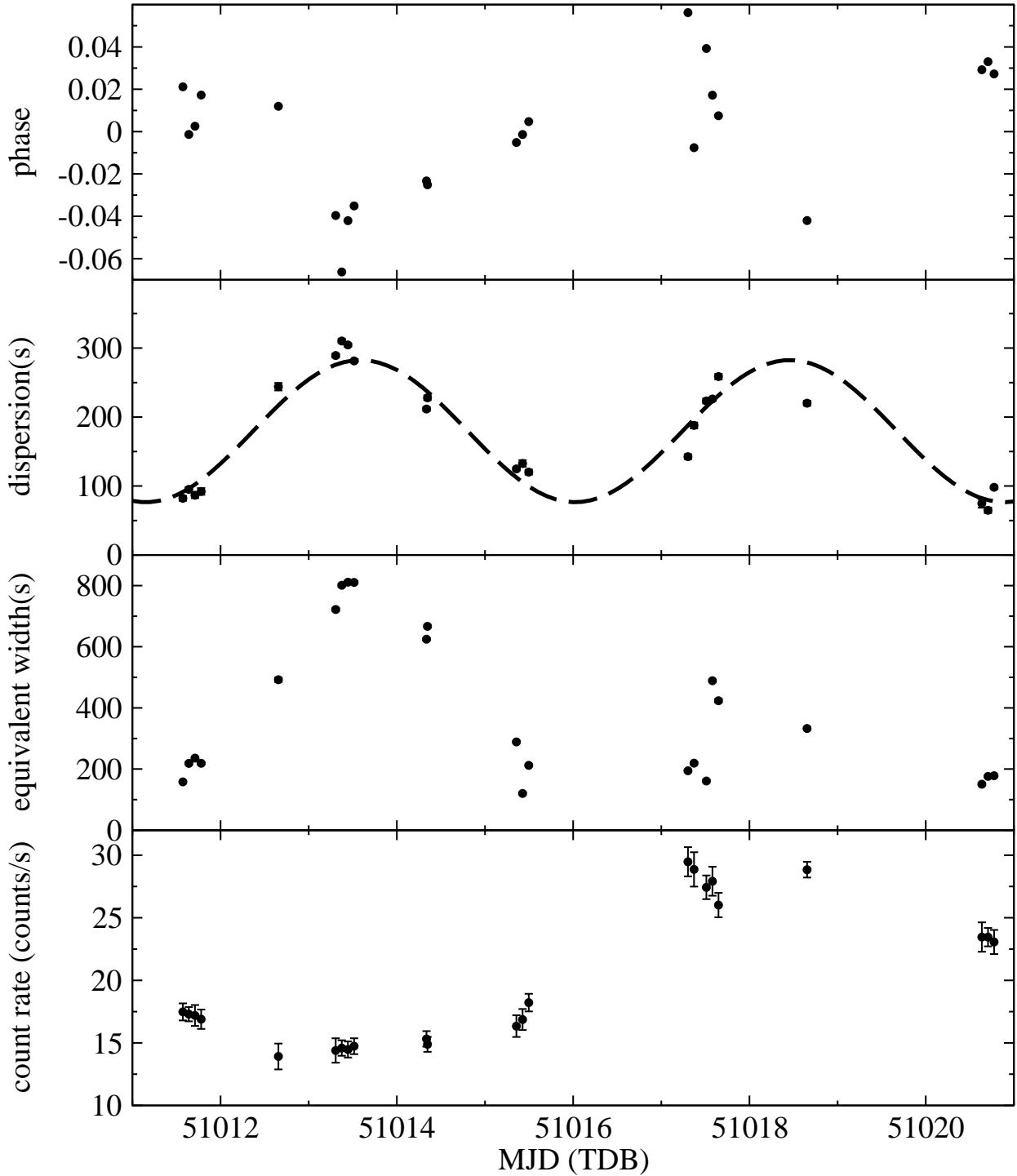} \caption{The dip parameters in 1998 10-day
observations. The most significant detection of periodicity is in
the dispersion. Through the sinusoidal fitting (dashed curve), the
period can be estimated as $(4.87 \pm 0.14)$ days.}
\label{1998_parameter}
\end{figure}

\begin{figure}
\plotone{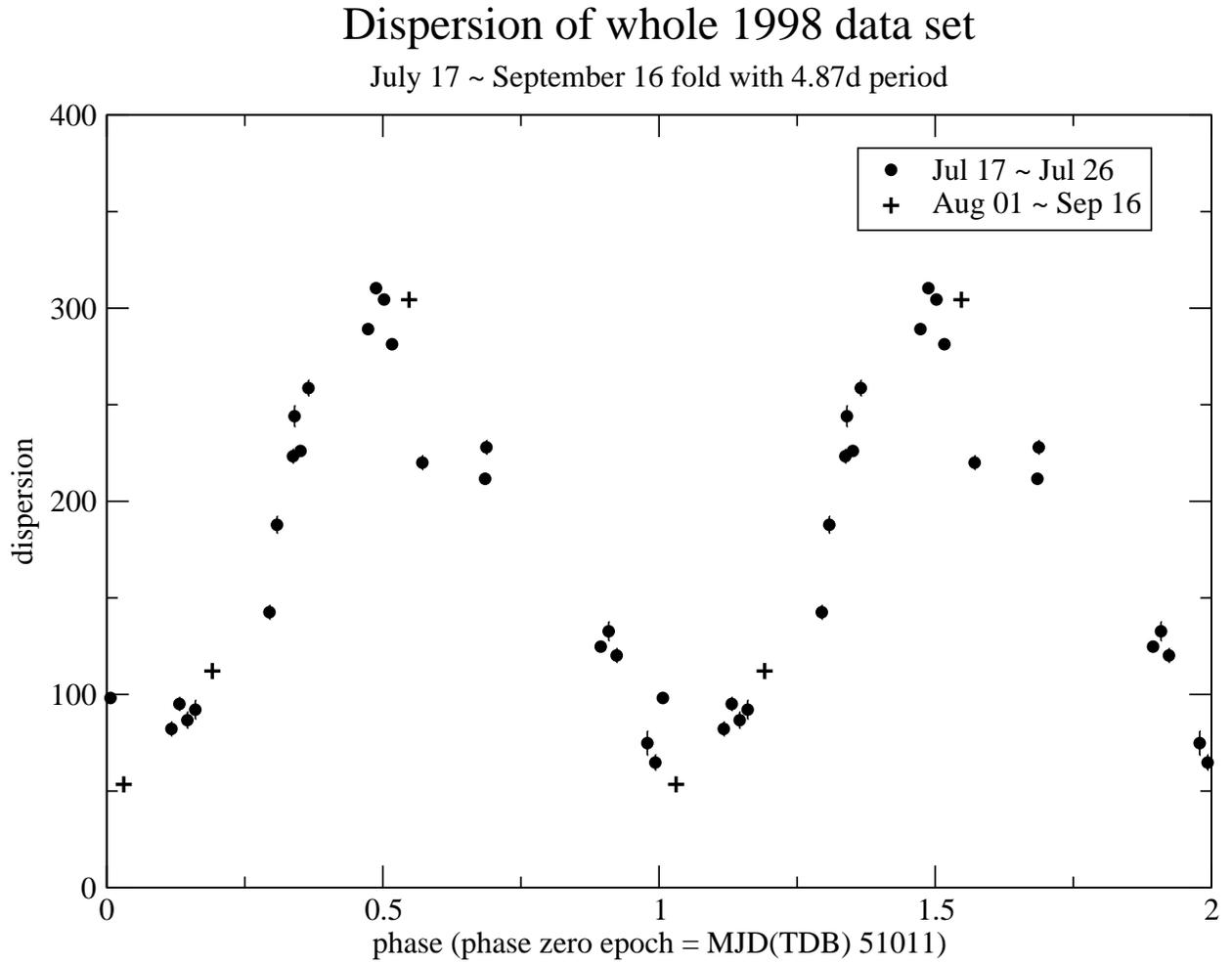} \caption{The dispersion of the complete 1998 data
folded with 4.87d period. The points on the diagram are 10-day
consecutive observation data, whereas the plus signs are other
observations in 1998.} \label{dispersion_fold}
\end{figure}

The new dip parameter definitions can be utilized to systematically study the variations of the parameters over a time scale of several days, which is owing to the probable apsidal precession or nodal precession period of the accretion disk.  Suitable data sets for searching periodicity are the 10-day consecutive RXTE observations in 1996 and 1998. For these two 10-day consecutive observation,  all dip parameters varied significantly over a time scale of 2-5 days (see Fig.
\ref{1996_parameter} and Fig. \ref{1998_parameter}).  Some of the parameters show strong periodic variations, especially for the dispersion in 1998 observation. The 1998 data dip dispersions exhibit a sinusoidal-like modulation with a period of (4.87 $\pm$ 0.14)d obtained by the sinusoidal fitting (see Fig. \ref{1998_parameter}).  This is consistent with the results on marginal dip phase variation proposed by \citet{homer2001}. We further folded the dispersions of the entire 1998 RXTE observation data with 4.87d period and found that all observed dip dispersions in 1998 are coherent (see Fig. \ref{dispersion_fold}).  It implies that the $\sim$4.87d modulation may stably last about two months. A $\sim$4d period variation in the dispersion, close to the $\sim$3.9d period variation proposed by \citet{chou2001}, can be seen in the 10-day consecutive observations in 1996, although the
periodicity is not as strong as the dispersions in the 1998 data (see Fig. \ref{1996_parameter}).

We also searched both data sets for the 3.9d and 4.87d beat side bands around the orbital period through the folding period search. This was provided by the FTOOLS subpackage \textit{efsearch}.  The 3.9d sidebands are clearly seen and the 4.87d sidebands are marginally detected in the 1996 observation. The detection of 4.87d sidebands is consistent with the negative superhump proposed by \citet{retter}.  Interestingly, no clear sideband was found for the 1998 data set although the dip dispersion shows such periodicity.

\subsection{Orbital Ephemeris}\label{oe}

We collected all available dip center times from the X-ray light curves observed from 1978 to 2002 and applied the linear ephemeris proposed by \citet{chou2001}:
\begin{equation}
T_{dipcenter}=MJD(TDB)50123.00944 \pm 1.4 \times
10^{-4}+\frac{3000.6508 \pm 0.0009}{86400} \times N
\end{equation}

\begin{figure}
\plotone{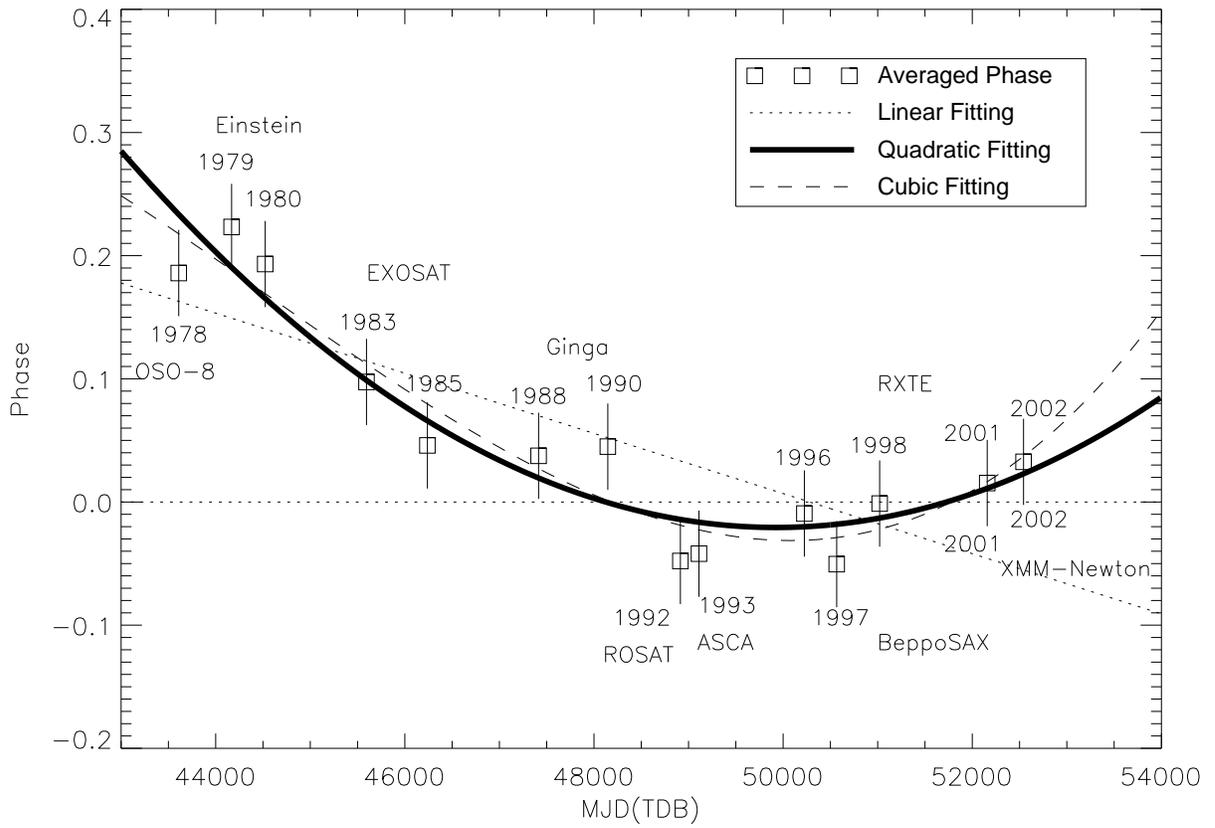} \caption{X-ray dip phase evolution of X1916-053
from 1978 to 2002. The folding period is 3000.6508s and the phase
zero epoch is MJD(TDB) 50123.00944. The thick curve is the
quadratic fitting result.} \label{phase_evolution}
\end{figure}

\noindent The dip phases are first re-binned yearly and a
$\pm$0.035 phase error was assigned for phase jitter (see
 \S \ref{std}). The long-term phase evolution is shown in Fig.
\ref{phase_evolution}. The linear ephemeris provided by
\citet{chou2001} is clearly no longer fit for 24 years of phase
evolution. The polynomial fittings for the dip phase and time
indicate that the quadratic fitting ($\chi^2_\nu = 0.75 $, dof=11)
is better than the linear fitting ($\chi^2_\nu = 2.61$, dof=12)
(see Fig. \ref{phase_evolution}) with a confidence level greater
than 99\% by the F-test. This implies that the orbital period
changed significantly with time from 1978 to 2002. No higher order
term is required for the dip phase evolution because the F-test
confidence level in comparison with cubic ($\chi^2_\nu = 0.72$,
dof=10) and quadratic fittings is only 74\%. We therefore updated
the orbital ephemeris in a quadratic form:
\begin{equation}
T_N=MJD(TDB)50123.00873\pm 0.0004+ \frac{3000.6511\pm
0.0007}{86400} \times N +(2.67 \pm 0.56)\times 10^{-13} \times N^2
\end{equation}

\noindent and the period derivative is:
\begin{equation}
\frac{\dot{P}_{orb}}{P_{orb}}=(1.62\pm 0.34) \times 10^{-7} yr^{-1}
\end{equation}
This positive value means that the orbital period increases with
time. \S \ref{cop} discusses further implications of orbital
change.

\subsection{The Stability of Dips}\label{std}

When the accretion stream impact the accretion disk, a bulge forms near the impact region.  However, the stream could penetrate the accretion disk, and forms a ring-like enhancement of the surface density (hereafter ``ring''). \citet{frank1987} proposed a model of two ``bulges.'' One lies on the impact region of the inflow stream at the accretion disk edge, at about 0.8-1.0 orbital phase relative to the companion star. Another bulge, make up with two-phase medium of cold clouds and hot intercloud gas forms after the point of accretion stream and ring, is lying on the disk ring between phase 0.3 to 0.8. This model successfully explains the dips for the eclipsing LMXB EXO 0748-676. Most dips of EXO 0749-676 concentrated around phases 0.65 and 0.9, which correspond to the two bulges of this model  \citep{frank1987}.  X1916-053 has no eclipse to configure the companion star position as a reference point so that it is hard to verify the bulge locations responsible for primary and secondary dips \citep{smale}. However, we believe that the bulge near the edge of the accretion disk is more stable than the inner bulge because it is closer to the companion star, and its location in the orbital frame is less affected by the accretion disk.

\begin{figure}
\plotone{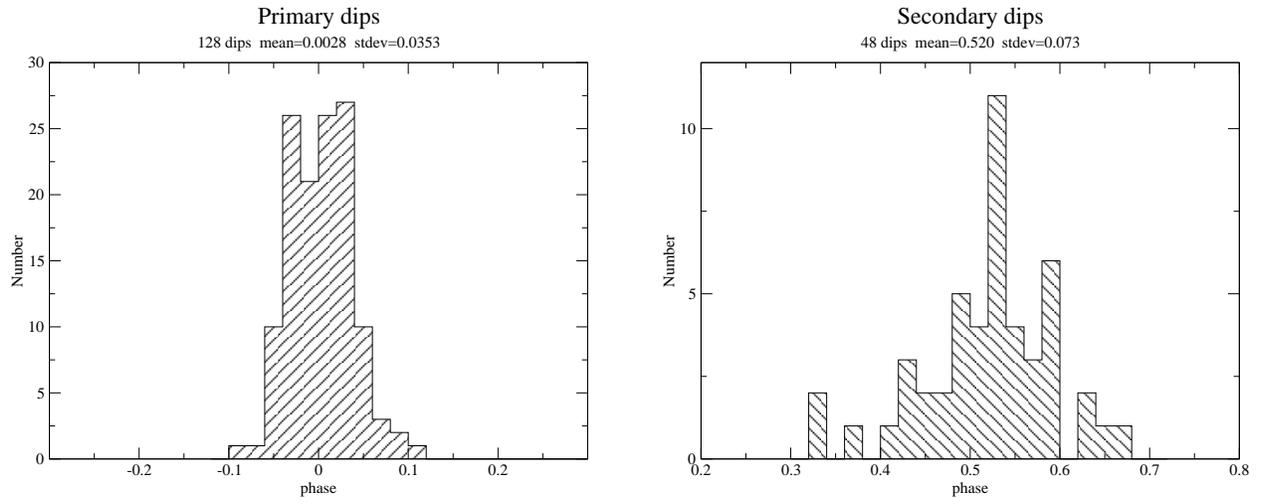} \caption{The phase distribution of primary dips
and secondary dips with respect to the quadratic ephemeris derived
from the primary dips. The standard deviation of primary dips is
significantly smaller than that of secondary dips. This means that
the primary dips are more stable than secondary dips.
  }
\label{dip_stat}
\end{figure}

To verify the bulge locations responsible for the primary and secondary dips, we tested the phase fluctuations (jitters) for both kinds of dips. The dip phases folded by the best quadratic ephemeris obtained by primary dips show that the standard deviation of the primary dips (0.035, 128 dips) is significantly smaller than the secondary dips (0.073, 48 dips) with a F-test null hypothesis probability of $6.3\times 10^{-11}$ (see Fig. \ref{dip_stat}).  However, since the quadratic ephemeris is yielded from minimizing the deviation of the primary dips (i.e. $\chi^2$ fitting), we also folded the dips with the best quadratic ephemeris yielded from the secondary dips to check consistency. The standard deviation of secondary dips (0.074) is still significantly larger than the primary dips (0.055), with a null hypothesis probability of 0.005. As a result, we conclude that the primary dip is due to the budge near the edge of accretion disk whereas the secondary dip is caused by the budge on the ring. The standard deviation of primary dips (0.035) with respect to the best ephemeris gives us an estimation of phase jitter. This estimation is used as the systematic error of the long-term phase evolution in \S \ref{oe}.

\section{Discussion}\label{discuss}

\subsection{Nodal Precession and Negative Superhump}

The negative superhump signal of nodal precession was detected by \citet{retter} in 1996 RXTE observations. They concluded that the signal owing to persistent state modulation rather than the dip width (or shape) variation because the signal still exists in the Lomb-Scargle power spectrum even when the dips were removed from the light curve.  In our analysis, although the periodicities of dip parameters are weak in 1996 RXTE observation, the periodicity of dip width variation in 1998 RXTE observation is strong.  This 4.87d period, confirmed in \S \ref{pp}, is consistent with Retter's result and provide another evidence of negative superhump. 

For a bulge lies on the outer edge of accretion disk, it is believed that the bulge will be more opaque and larger width near the disk plane.  Once while the accretion disk has retrograde nodal precession, the angle between disk plane and observer's line of sight varies with nodal precession period (see Fig. 6.18 in \citet{hellier2001}).  As a result, dip width variation can be interpreted easily by the variation of the angle of disk plane. The 4.74d dip phase variation proposed by \citet{homer2001} may be induced by the dip width variation due to asymmetric bulge geometry.

The mass ratio of X1916-053 can be interpreted by the negative
superhump period.  \citet{chou2001} derived its mass ratio of  0.022 or 0.011 from its 3.9d positive superhump or a Roche-lobe filled white dwarf respectively.  On the other hand, for the negative superhump, \citet{larwood} proposed that the relation between the nodal precession period of the disk and the particle frequency of the outer accretion disk can be represented as:
\begin{equation}
\frac{\omega_n}{\Omega(R)}=-\frac{15}{32}qr^3\cos\delta
\end{equation}
where $\omega_n$ is the frequency of nodal precession, $\delta$ is the tilted angle of accretion disk, $\Omega(R)$ is the particle frequency of outer accretion disk, and a ratio of specific heats $\gamma=5/3$ is assumed.  For $\Omega(R)=3\omega_{orb}$ and assuming a small $\delta$, the mass ratio can be estimated as $q \sim 0.045$, slightly larger than the ones proposed by \citet{chou2001}.  All the models imply that
the companion's mass is less than the lower limit of a normal
main-sequence star ($\lesssim 0.08 M_{\odot}$), consistent with the previous prediction that the secondary must be a fully-degenerated or semi-degenerated white dwarf.

\subsection{Orbital Period Change}\label{cop}

Table \ref{period_change} lists the five LMXBs with significant detection of orbital period changes prior to this research. Four
of them (EXO 0748-676, 4U 1820-30, X1822-371 and Cyg X-3) have
period derivatives inconsistent with the standard model proposed
by \citet{rappaport1987} \citep[see][]{tavani1991}. The standard model, in which mass loss and orbital period change due to gravitational radiation, predicted a positive orbital period derivative for the LMXBs with degenerate companions. X 1916-053 is a LMXB composed of a neutron star and a white dwarf. Using the model proposed by \citet{rappaport1987},  the orbital period derivative would be  $\dot{P}_{orb}/P_{orb}=5.96 \times 10^{-10} yr^{-1}$, which is a factor of $10^2$ to $10^3$ smaller than the observed value. The period derivative value predicted by this model is consistent with the observed value only when the mass transfer is extremely not conservative (near the singularity in this model).

\citet{tavani1991} proposed a radiation-driven model to explain inconsistencies between the standard model and the observed value.  They argued that the companion star may be illuminated by a relatively large flux of radiation from the primary star or accretion disk. Such radiation can drive a strong evaporative wind from its outer atmosphere. In this model, the irradiated companion can transfer mass even if it does not fully fill its Roche lobe.  They derived an average self-sustained value of radiation-driven mass loss rate of $-\dot{m}_2\approx 10^{-8} M_{\odot} yr^{-1}$, which is much larger than that of gravitational radiation ($-\dot{m}_{GR}\approx 9.15\times 10^{-12}M_{\odot}yr^{-1}$).  For X 1916-053, the period derivative value can be interpreted by this model with $60-90$ \% of companion mass loss outflow from the system. The mass outflow has been detected in LMXBs.  \citet{chakrabarty2003} reported that a strong He I 1.083 $\mu m$ emission line with P Cygni profile in GX 1+4/V2116 Oph through infrared observation. From the blue edge of this profile, they inferred that there is an outflow with a velocity much faster than a typical red giant wind from this binary system.  For X1916-053, its outflow could be verified if the P Cygni profile were detected in its optical counterpart.  Unfortunately, the optical counterpart of X1916-053 is too dim (V=21) so that no optical spectroscopy has ever been reported.  From the mass accreted onto the neutron star predicted by \citet{tavani1991}, the X-ray luminosity can be estimated as $1-4 \times 10^{37} erg\cdot s^{-1}$. This is slightly larger than the observed value ($0.5-1.44 \times 10^{37} erg\cdot s^{-1}$, \citet{bloser2000}).

This research has made use of data obtained from the High Energy Astrophysics Science Archive Research Center (HEASARC), provided by NASA's Goddard Space Flight Center. This research is partially supported by grant NSC 93-2112-M-008-007 and NSC 94-2112-M-008-003 of the National Science Council.

\clearpage

\input{tab1.tex}
\input{tab2.tex}
\input{tab3.tex}
\input{tab4.tex}

\end{document}

%% file: tab1.tex
\begin{table}
\caption{The historical X-ray observations of X1916-053}
\label{obs_status}
\begin{center}
\begin{tabular}{llll}
\tableline\tableline
Year & observatory & observation date & number of dips \tablenotemark{1}\\
\tableline
2002 & XMM-Newton & Sep. 25\tablenotemark{2} \\
 & RXTE & Sep. 25 & 7\\
 2001 & RXTE & May 25, Jun 16, 17, 30, Jul 01, Oct 01 & \\
  & BeppoSAX & Oct 01-02 & 27\\
 1998 & RXTE & Jun 24, Jul 17-26\tablenotemark{3}, Aug 01, 10, Sep 14, 16 & 26\\
 1997 & BeppoSAX & Apr. 27-28 & 9\\
 1996 & RXTE & Feb 10, Mar 13, May 05, 14-23\tablenotemark{3}, Jun 01,  & \\
  & & Jul. 15, Aug 16, Sep 06, Oct 29 & 27\\
1993 & ASCA & May 02-03 & 5\\
1992 & ROSAT & Oct 17-19 & 1\\
1990 & Ginga & Sep 11-13 & 2\\
1988 & Ginga & Sep. 09-12 & 2\\
1985 & EXOSAT & May 24, Oct 13 & 10\\
1983 & EXOSAT & Sep 17 & 8\\
1980 & Einstein & Oct 11 & 2\\
1979 & Einstein & Oct 22 & 1\\
1978 & OSO-8 & Apr 07-14 & 1\\
\end{tabular}

\tablenotetext{1}{This column represents number of dips of this year.  However, some light curves contain no complete dips or have relative lower significance are folded into one fold light curve so that we treat them as one dip per fold light curve.  }

\tablenotetext{2}{The simultaneous observation of RXTE and
XMM-Newton} 
\tablenotetext{3}{The 10-day consecutive observations
in 1998 and 1996}

\end{center}
\end{table}

%% file: tab2.tex
\begin{table}
\centering \caption{Dip parameters vs energy bands.  The dip
center time is measured from MJD(TDB) 52542}
\label{parameter_band}
\begin{tabular}{ccccc}
\tableline\tableline
dip number &Energy band (keV)& dip center time (s) & dispersion (s) & equivalent width (s)\\
\tableline
1 &0.7-8.0 (total)& 15619.323 & 193.476 & 444.308\\
  &0.7-1.7 (soft) & 15624.551 & 194.775 & 485.480\\
  &1.7-8.0 (hard) & 15616.351 & 192.344 & 431.547\\
\tableline
2 &0.7-8.0 (total)& 18767.322 & 187.421 & 291.581\\
  &0.7-1.7 (soft) & 18766.314 & 191.611 & 341.541\\
  &1.7-8.0 (hard) & 18769.707 & 183.034 & 250.401\\
\tableline
3 &0.7-8.0 (total)& 21771.504 & 223.684 & 415.795\\
  &0.7-1.7 (soft) & 21776.967 & 220.304 & 471.319\\
  &1.7-8.0 (hard) & 21765.254 & 227.488 & 370.354\\
\tableline
4 &0.7-8.0 (total)& 24769.473 & 243.336 & 474.337\\
  &0.7-1.7 (soft) & 24771.145 & 245.808 & 541.436\\
  &1.7-8.0 (hard) & 24769.922 & 239.112 & 417.391\\
\tableline
5 &0.7-8.0 (total)& 27723.248 & 250.468 & 457.142\\
  &0.7-1.7 (soft) & 27729.359 & 249.732 & 516.507\\
  &1.7-8.0 (hard) & 27716.568 & 249.946 & 404.874\\
\tableline
\end{tabular}
\end{table}

%% file: tab3.tex
\begin{table}
\centering \caption{Measured dip center time between XMM-Newton MOS detector and RXTE PCA}
\label{dct_satellite}
\begin{tabular}{rcc}
\tableline\tableline
dip number & dip center time(s) & dispersion (s) \\
\tableline
1(XMM-Newton) & 15619.323 & 193.48\\
(RXTE) & 15603.40 & 182.708\\
\tableline
2(XMM-Newton) & 21771.504 & 223.684\\
(RXTE) & 21730.94 & 225.89\\
\tableline
3 (XMM-Newton) & 27723.248 & 250.468\\
(RXTE) & 27685.86 & 254.65\\
\tableline
\end{tabular}
\end{table}

%% file: tab4.tex
\begin{table}
\caption{LMXBs characteristics with measured orbital period
change} \label{period_change}
\begin{center}
\begin{tabular}{lllll}
\tableline\tableline
Object Name & modulation \tablenotemark{*} &$P_{orb}$ (hr) & $\dot{P}_{orb}/P_{orb}$ (yr$^{-1}$) & Reference\\
\tableline
4U 1820-30 & M & 0.19 & $-3.74 \times 10^{-8}$ & \citet{chou_grindlay}\\
X 1916-053& D & 0.83 & $1.62 \times 10^{-7}$ & this work\\
EXO 0748-676 \tablenotemark{1}& D,E & 3.82 & $2.7 \times 10^{-8}$ & \citet{wolff2002}\\
Cyg X-3 \tablenotemark{2}& M & 4.82 & $1.05 \times 10^{-6}$ & \citet{singh2002}\\
X 1822-371& PE,D & 5.57 & $3.4 \times 10^{-7}$ & \citet{hellier1990}\\
Her X-1 & E & 40.8 & $-1.32 \times 10^{-8}$ & \citet{deeter1991}\\
\end{tabular}

\tablenotetext{*}{E: total eclipse, PE: partial eclipse, D:
periodic dips, M: other modulation}
\tablenotetext{1}{The period
derivative value of EXO 0748-676 came from the quadratic fitting of entire dataset
from EXOSAT(1985) to RXTE(2000).}
\tablenotetext{2}{Cyg X-3 was
identified to be a High Mass X-ray binary because the companion
star was confirmed as a Wolf-Rayet helium star with a strong
wind.}

\end{center}
\end{table}